# Computational framework for polymer synthesis to study dielectric properties using polarizable reactive molecular dynamics

Ankit Mishra[1], Lihua Chen[2], ZongZe Li[3], Ken-ichi Nomura[1], Aravind Krishnamoorthy[1], Shogo Fukushima[5], Subodh C. Tiwari[1], Rajiv K. Kalia[1], Aiichiro Nakano[1], Rampi Ramprasad[2], Greg Sotzing[4], Yang Cao[3], Fuyuki Shimojo[5] and Priya Vashishta[1]

[1]Collaboratory for Advanced Computing and Simulations, University of Southern California, Los Angeles, CA 90089-0242,
[2]School of Materials Science and Engineering, Georgia Institute of Technology, Atlanta, GA 30332,
[3]Department of Electrical and Computer Engineering, [4]Department of Chemistry, University of Connecticut, Storrs, CT 06269
[5]Department of Physics, Kumamoto University, Kumamoto, 860-8555, Japan

**Abstract**
The increased energy and power density required in modern electronics poses a challenge for designing new dielectric polymer materials with high energy density while maintaining low loss at high applied electric fields. Recently, an advanced computational screening method coupled with hierarchical modelling has accelerated the identification of promising high energy density materials. It is well known that the dielectric response of polymeric materials is largely influenced by their phases and local heterogeneous structures as well as operational temperature. Such inputs are crucial to accelerate the design and discovery of potential polymer candidates. However, an efficient computational framework to probe temperature dependence of the dielectric properties of polymers, while incorporating effects controlled by their morphology is still lacking. In this paper, we propose a scalable computational framework based on reactive molecular dynamics with a valence-state aware polarizable charge model, which is capable of handling practically relevant polymer morphologies and simultaneously provide near-quantum accuracy in estimating dielectric properties of various polymer systems. We demonstrate the predictive power of our framework on high energy density polymer systems recently identified through rational experimental-theoretical co-design. Our scalable and automated framework may be used for high-throughput theoretical screenings of combinatorial large design space to identify next-generation high energy density polymer materials.

## 1. Introduction:

Polymers are an important class of materials owing to their chemical resistance, light weight, high processability and ability to act as thermal and electrical insulators. Because of their useful properties, polymer products are ubiquitous in our daily lives and are used in plastics, packaging, military, aerospace and various other fields.[1] In insulation, isolation and electrostatic energy storage technologies, polymers dielectric materials have become mainstays, since they are relatively inexpensive and lighter than ceramic material alternatives. For example, biaxially-oriented polypropylene (BOPP) is currently employed as state-of-the-art polymer dielectric films in electrostatic energy storage capacitors. Despite having high dielectric strength and low dielectric losses, BOPP exhibits relatively low dielectric constant and energy density.[2-4] Therefore, concerted experimental and theoretical efforts have been made to develop new class of dielectric polymer materials which possess low conduction losses under high electric fields and electrical energy density and can be used in a wide range of applications in modern electronics and electric power systems.[5-7]

Due to the rapid advances in machine-learning technologies, the closed-loop cycle consisting of materials synthesis, characterization and computational modeling has been attracting great attention to accelerate design and discovery of novel materials.[8-15] Advanced computational screening procedures also have been applied and have significantly helped identify important high energy density polymeric materials.[16, 17] Since polymers exist as repeating monomer units, it is possible to express them as a series of strings and store in relational databases.[18] The database can be utilized to create models using machine learning tools to predict relevant polymer properties.[19] This idea has enabled fast computational screening, coupled with hierarchical modelling, to accelerate the identification of promising repeat units to synthesize.[16, 20] For instance, polyurea and polythiourea, with -NH-CO-NH-$C_6H_4$- and -NH-CS-NH-$C_6H_4$- repeat units, have been identified as high dielectric constant organic polymers utilizing the rational co-design strategy.[16, 21] The identified repeat units have been used as starting points for further optimization of the chemistry so as to be suitable for small- and large-scale thin film processing.[21] One example of such an optimized polymer is PDTC-HK511, a polythiourea-based material by polymerizing p-phenylene isothiocyanate and an etheramine jeffamine HK511.[22] Similar protocols have been used to create TDI-EDR148, a polyurea-based material by addition polymerization of 1,4-toluene diisocyanate (TDI) and an etheramine jeffamine EDR-148. These materials have been further characterized and studied for establishing a relationship between their microscopic molecular structure and macroscopic dielectric properties.[16, 22, 23] They have been found to have high breakdown strength, energy density, dielectric constant ($> 5$) and maintain low losses at high electric fields. Moreover, this advanced screening procedure has identified promising high energy density polymers from the existing materials such as divinyl siloxane benzo-cyclobutene (DVS-BCB) and flurorene polyester (FPE).[24, 25] DVS-BCB and FPE exhibit high glass transition temperatures ($T_g > 300$ °C) and high dielectric constants ($> 2.5$) at room temperature.

To further accelerate the computational-screening step, it is crucial to develop new descriptors that capture the effects of complex chemistries and microstructures on the temperature-dependent dielectric behavior while retaining quantum-mechanical (QM) accuracy. First-principles molecular dynamics (MD) simulation[26-28] is capable of describing distinct polymer chain structures that consist of designed repeat units, although the computational cost prohibits to investigate complex local structures such as domains and interfaces between crystalline, amorphous, and semi-crystalline phases. It has been shown that crystalline polymer materials exhibit lower dielectric constant values due to restrictive dipolar relaxations caused by the surrounding media. On the other hand, significantly higher dielectric constant values have been observed in amorphous phases, in which the vast difference has been attributed to the increased spatial distributions in the amorphous samples.[29] Critical role of morphology has further been investigated for many polyolefins such as polyethylene and polypropylene.[29, 30] However, system sizes handled by the first-principles approach is limited to a few hundred atoms due to the high computational cost.

All-atom MD simulations using an empirical force field can be used to scale the simulation system size up to practically relevant polymer lengths. Such a force field can accurately describe structural motifs and features such as repeat units, single chain morphology, chain stackings, local domains and their interface, as well as temperature dependence and thermo-mechanical properties. However, due to the absence of the polarizability of the constituent materials as a function of their valence charge state[31] this approach suffers from poor predictions of dielectric response subjected to an electric field.

To circumvent this deficiency, we have developed a scalable computational framework based on a charge-state aware ReaxPQ+ model. The original ReaxPQ model combines all-atom reactive molecular dynamics (RMD) employing reactive force field (ReaxFF)[32] and a recently-proposed polarizable charge model (PQEq)[33]. To study dielectric response under electric field, ReaxPQ+ model was introduced to incorporate the effects from both internal and external electric fields. ReaxPQ+ has been successfully applied to a number of inorganic and organic materials[34], however, the absence of charge-state in the ReaxPQ+ model resulted in time-consuming and laborious parameter tunings, presenting a serious challenge to enable a fully automated high-throughput computational screening for polymer dielectrics. To this goal, we here introduce a charge-state aware ReaxPQ+ model to accurately describe the dielectric responses of amorphous polymers based on the quantum-mechanically informed atomic polarizability. Once model parameters have been developed, our scalable framework predicts dielectric properties of polymers within a small fraction of QM calculation time[34] and handles industrially-relevant polymer chain lengths in a highly-automated fashion. We have applied our model to computationally synthesize amorphous polymer systems, and demonstrated the role of morphological complexity in evaluating the dielectric constant of polymers.[29, 30] In addition, we successfully apply the framework validated by the dielectric responses of PDTC-HK511 and TDI-EDR148 to newly identified FPE and DVS-BCB within an experimental accuracy. The subsequent sections present the key components of the framework, obtained dielectric properties and discussions.

## 2. Methodology

Our framework can be divided into following components: a) synthesis of polymer system, and b) ReaxPQ+ model development and dielectric-constant estimate (Fig. 1). Details of these components are described in subsequent subsections.

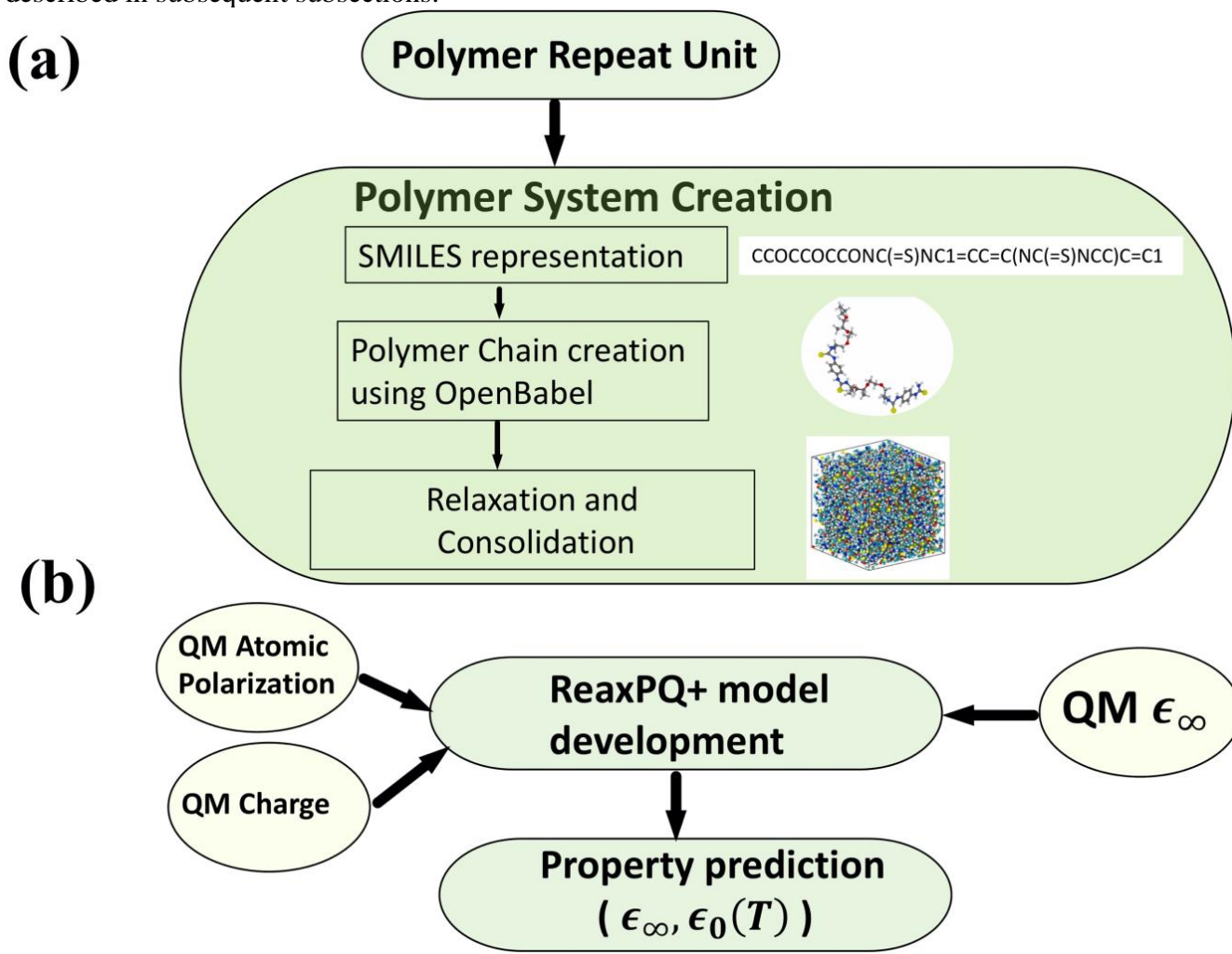

**Figure 1:** Computational framework of polymer dielectric-behavior estimate. (a) Synthesis of amorphous polymer system. SMILES string and Open Babel are used to create a single polymer chain. Multiple

polymer chains are placed at sufficiently large distance from each chain in a simulation system. The simulation system is subjected to a number of consolidation and relaxation steps until the system reaches to a desired density. (b) ReaxPQ+ model development that involves various QM-based calculations and validations such as atomic polarization and charges for constituent atomic species, and dielectric constant estimate.

## 2.1. Charge-State Aware ReaxPQ+ Model

To take into account the dielectric response to both internal and external electric fields, the ReaxPQ+ model employs the polarizable charge-equilibration scheme called PQEq,[33] which in turn introduce core and shell charges to capture the complex interplay between the electronic and ionic dynamics. The potential energy $E(\{\mathbf{r}_{ij}\},\{\mathbf{r}_{ijk}\},\{\mathbf{r}_{ijkl}\},\{q_i\},\{BO_{ij}\})$ of the system in ReaxPQ+ is represented as a function of relative positions of atomic pairs $\mathbf{r}_{ij}$, triplets $\mathbf{r}_{ijk}$ and quadruplets $\mathbf{r}_{ijkl}$, atomic charges $q_i$ and bond orders $BO_{ij}$ between different atomic pairs. In ReaxPQ+, the total charge on $i$-th atom is computed based on the contribution from a massless shell ($\rho_{is}$) connected to the core ($\rho_i$) by an isotropic harmonic spring constant ($k_s$) as shown in Fig. 2. The massless shell ($\rho_{is}$) has fixed total charge $-Z_i$ while the core ($\rho_i$) charge consists of two parts: 1) a dynamically variable atomic charge ($q_i$), and 2) a fixed charge ($Z_i$) compensating the massless shell counterpart.

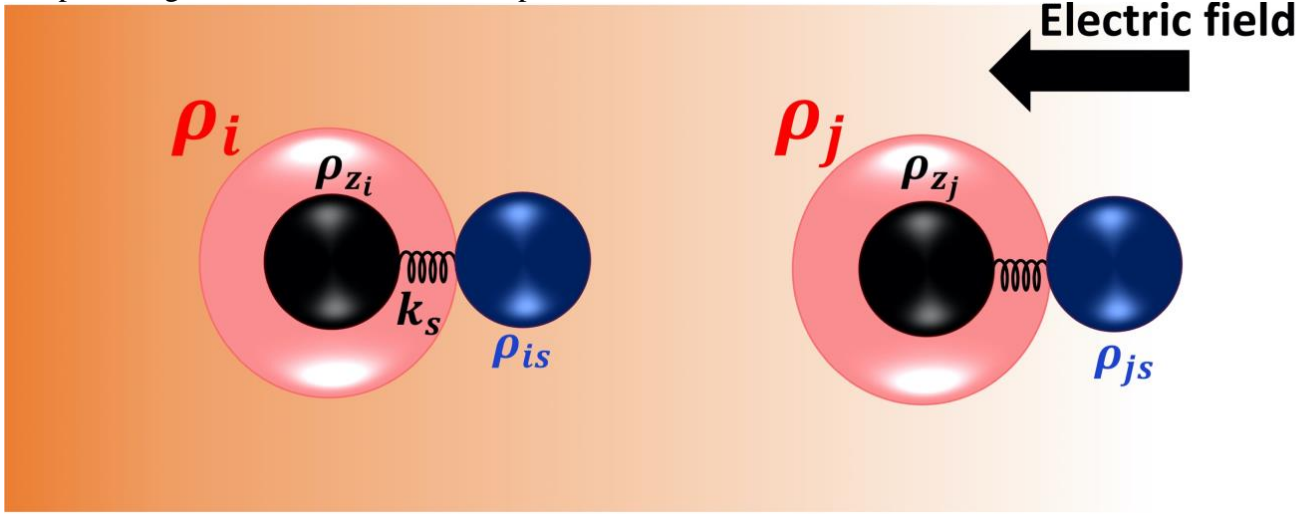

**Figure 2**: Schematic of the response of core (black) and shell (red) charges to an external electric field in ReaxPQ+ model.

The atomic charges $q_i$ are determined by minimizing the total Coulomb energy $E_{Coulomb}$ at every step until the electrochemical potentials, $\partial E_{Coulomb}/\partial q_i$, become equal among all atoms. The Coulombic energy is described as

$$E_{Coulomb}(\{\vec{r}_{ic}, \vec{r}_{is}, q_i\})$$
$$= \sum_i^N \left\{E_i^0 + \chi_i^0 q_i + \frac{1}{2}J_{ii}^0 q_i^2 + \frac{1}{2}K_s r_{ic,is}^2\right\}$$
$$+ \sum_{i>j}\{C(\vec{r}_{ic,jc})q_{ic}q_{jc} - C(\vec{r}_{ic,js})q_{ic}Z_j - C(\vec{r}_{is,jc})q_{jc}Z_i + C(\vec{r}_{is,js})Z_iZ_j\}, \quad (1)$$

where $\vec{r}_{ic}, \vec{r}_{is}, \chi_i^0$ and $J_{ii}^0$ are the positions of core and shell charges, the electronegativity and the hardness of $i$-th atom, respectively. $r_{ia,jb}$ ($i, j = 1,…, N$; $a, b$ = core(c) or shell(s)) represents the charge-charge distances. The electrostatic energy between two Gaussian charges is given in terms of the error function $C_{ia,jb}(r_{ia,jb})$, and the Coulombic interaction is screened using a taper function $T(r)$.[33] The shell positions $r_{is}$ subjected to an external electric field is determined

$$F_{inter} + F_{external} = F_{intra} \quad (2)$$

$$\mathbf{F}_{inter} = -\frac{\partial}{\partial \mathbf{r}_{is}}\left\{\sum_{ia>jb} T(r_{ia,jb})C_{ia,jb}(r_{ia,jb})q_{ia}q_{jb}\right\},$$

$$\mathbf{F}_{intra} = -\frac{\partial}{\partial \mathbf{r}_{is}}\left(\frac{1}{2}K_s r_{ic,is}^2\right) \text{ and } F_{external} = \sum_{ia} q_{ia}\varepsilon.$$

ReaxPQ+ model describes the system polarization *via* the displacement of Gaussian-shaped shell-charge relative to the core-charge position, which is controlled by two key parameters: the core-charge radius $R_c$ and the spring constant $K_s$ between the core-shell charges. A critical component of the ReaxPQ+ model is the choice of atomic polarizability dictated by $K_s$, which substantially changes depending on their valence charge state. Therefore, the model must be aware of the atomic state to realize accurate prediction of dielectric response as a system. For determining the polarizability of constituent atomic species, we use Hartree-Fock calculations with def2-svpd basis sets[35]. Table 1 summarizes the obtained atomic polarizability values with various charge states. The polarizability of neutral atoms is presented as validation in the right-most column, which shows good agreement with values in literature[36]. These polarizability values can be used for initial estimation of $K_s$:[33]

$$K_s = C\frac{Z_i}{\alpha}, \quad (3)$$

where $C$ is a constant conversion factor, $Z_i$ is the charge on shell attached to core, and $\alpha$ is the atomic polarizability (Å$^3$), as reported in Table 1.

In addition to the polarizability, an accurate description of atomic charges may improve the estimation of dielectric constants. In this study, we use QM charges throughout our simulations for prediction of dielectric properties. Table S1 in the supporting information presents optimized ReaxPQ+ parameters for the constituent atoms, which are validated using the high-frequency dielectric constant $\epsilon_\infty$ obtained by Berry-phase method[37, 38].

| charge state | +2 | +1 | 0 | -1 | -2 | literature[36] |
|---|---|---|---|---|---|---|
| C | 0.70 | 0.97 | 1.70 | 3.07 | NA | 1.71 |
| N | 0.41 | 0.61 | 0.99 | 1.59 | NA | 1.10 |
| O | 0.27 | 0.40 | 0.69 | 1.81 | 4.12 | 0.80 |
| S | 1.09 | 1.60 | 2.77 | 5.77 | 11.40 | 2.90 |
| H | NA | NA | NA | 2.45 | 2.651 | 0.670 |

**Table 1:** Computed atomic polarizability (Å$^3$) for various charge states of constituent atom species.

### 2.2. Amorphous Polymer Model Creation

The three major steps in the amorphous polymer model creation using TDI-EDR148 as an example system is illustrated in Fig. 3. The initial step involves generating a Simplified Molecular Input Line Entry System (SMILES) string for a polymer structure under consideration. SMILES is a specification in the form of a line notation for describing the structure of chemical species using short ASCII strings.[39] In the second step, Open Babel[40] is used to generate a low-energy structure based on the input SMILES string.

Open Babel is a computer software, a chemical expert system mainly used to convert chemical file formats. The obtained polymer-chain structure is relaxed using conjugate-gradient method, followed by RMD simulation at low temperature to relax the bonds, angles, dihedrals and monomer linkages connecting different repeat units together. The final step involves stacking the polymer chains, a series of RMD simulations to consolidate and relax the system to achieve a desired density at a given temperature. Fig. S1 shows amorphous structures produced for PDTC-HK511, FPE and DVS-BCB.

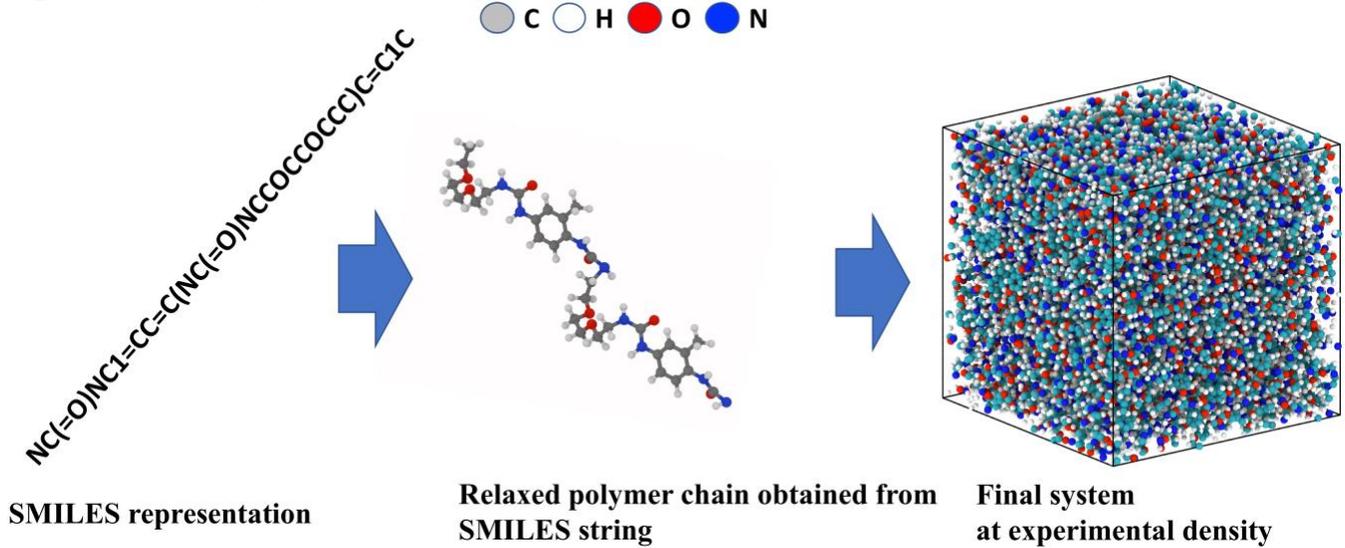

**Figure 3:** Amorphous-polymer synthesis of TDI-EDR148 involves accurate SMILES representation of the polymer followed by its generation using Open Babel and final consolidation to achieve desired density.

Scalable polymer synthesis is realized using a parallel MD engine called RXMD[41]. RXMD employs first principles-informed ReaxFF[32] and a divide-and-conquer strategy to achieve the scalability from desktop computer to supercomputing platforms (Fig. S2). Message Passing Interface (MPI) library is used in data exchanges between processes and inter-process communications[42]. RXMD supports ReaxPQ+[33] model, thus allowing accurate study of the polarization and dielectric response of materials subjected an electric field.

**2.3. Estimation of Dielectric Constant**

The temperature-dependent dielectric constant $\epsilon_0(T)$ is calculated using Eq. (4), which consists of the high-frequency dielectric constant ($\epsilon_\infty$) and the temperature-dependent term,

$$\epsilon_0 - \epsilon_\infty = \frac{\langle \Delta M^2 \rangle}{\epsilon_{vac} 3\Omega k_B T}, \qquad (4)$$

where $\Omega$ is the volume of cell, $k_B$ is the Boltzmann constant, $T$ is the temperature, and $\langle \Delta M^2 \rangle \equiv \langle M^2 \rangle - \langle M \rangle^2$ is the variance of the total dipole moment $\boldsymbol{M}(t)$ over time $t$. Each component of the total dipole moment is computed as follows:

$$\mathbf{M}(t) = \left(M_x(t), M_y(t), M_z(t)\right) = \sum_j^{N_{mol}} \sum_\alpha^{N_{atom}} [\mathbf{r}_{\alpha j}(t) - \mathbf{r}_{0j}(t)] q_{\alpha j} \qquad (5)$$

$$M_x(t) = \sum_j^{N_{mol}} \sum_\alpha^{N_{atom}} [x_{\alpha j}(t) - x_{0j(t)}] q_{\alpha j} \qquad (5a)$$

$$M_y(t) = \sum_j^{N_{mol}} \sum_\alpha^{N_{atom}} [y_{\alpha j}(t) - y_{0j(t)}] q_{\alpha j} \qquad (5b)$$

$$M_z(t) = \sum_j^{N_{mol}} \sum_\alpha^{N_{atom}} [z_{\alpha j}(t) - z_{0j(t)}] q_{\alpha j} \qquad (5c)$$

Here, $N_{atom}$ is the number of atoms in molecule $j$, $q_{\alpha j}$ is the charge of atom $\alpha$, $\mathbf{r}_{\alpha j}(t)$ is the position vector of atom $\alpha$ with components $x_{\alpha j}(t), y_{\alpha j}(t), z_{\alpha j}(t)$ and $r_{0j}(t)$ is the position of a reference atom in molecule $j$ with coordinates $x_{0j}(t), y_{0j}(t), z_{0j}(t)$. The high-frequency dielectric constant $\epsilon_\infty$ is obtained from the polarization due to the shell charge displacement at fixed atom positions subjected an electric field. We obtain the best estimates for $\epsilon_\infty$ by further tuning the initial parameter estimates of the ReaxPQ+ model to fit the first-principles Berry-phase method.[38, 43, 44] Once the model has been developed, the framework described in this section allows researchers to investigate a large sets of repeat unit structures and quickly evaluate the morphology effect such as crystal versus amorphous and system temperature in their dielectric properties. These insights are crucial to develop new descriptors to further accelerate the theoretical screening process for novel high energy density polymer design. Section S3, S4, and S5 in supporting information present the effects of polymer morphology and temperature, and the high-frequency dielectric constants of various model polymer structures, respectively.

## 3. Result and Discussion

In this section, we demonstrate our framework on $\epsilon_0(T)$ of known high density polymers (PDTC-HK511 and TDI-EDR148) and newly identified polymers (DVS-BCB and FPE[24, 25]). All amorphous systems are created at experimental density[16, 22] following the procedures discussed in Section 2.2.

### 3.1. Dielectric Behavior of TDI-EDR148 and PDTC-HK511

Fig. 4 (a) and (b) show the fluctuation in the total dipole moment $\mathbf{M}(t)$ of the amorphous TDI-EDR148 system at temperatures 100 and 300 K, respectively. We have used a large number of independent RMD trajectories (see Section S4 in supporting information for details) to ensure good statistics without the need of very long-time trajectories. Eq. (5) indicates that the dielectric constant is proportional to the ratio between the variance in the total dipole moment and the system temperature. We observe that the increase of the total dipole moment variance surpasses the temperature term in the amorphous TDI-EDR148 system. We confirm the same trend at various temperatures as well as in amorphous PDTC-HK511 system, see Figs. S3 and S4.

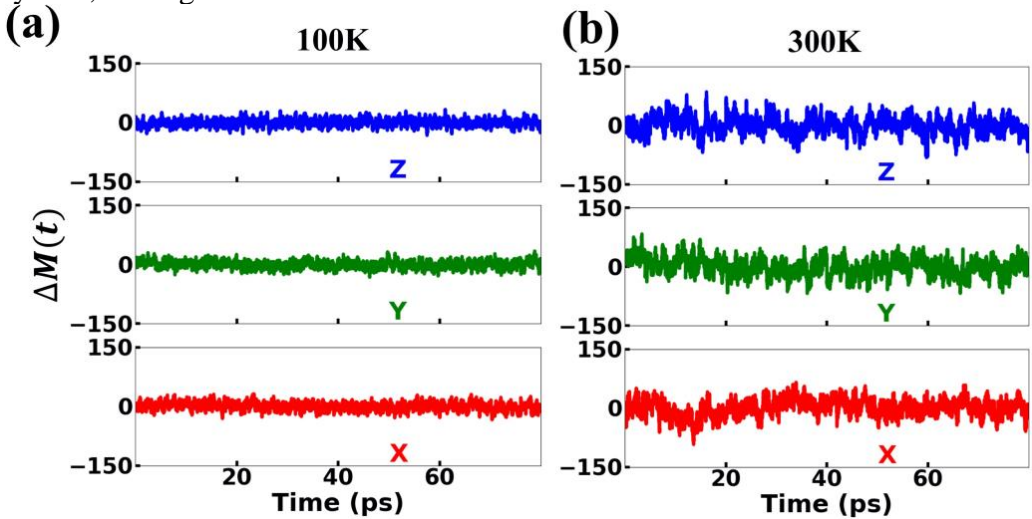

**Figure 4:** Time evolution of dipole moment in TDI-EDR148 at (a) 100 K and (b) 300 K. The dipole moment (measured in units of Debye) variation in x, y and z direction are shown in red, green and blue color, respectively. The increase in the dipole moment fluctuation with respect to temperature is manifested in dielectric constant.

Fig. 5 (a) and (b) show temperature dependence of the dielectric constants of PDTC-HK511 and TDI-EDR148, respectively. The high-frequency dielectric constant $\epsilon_\infty$ is estimated by ReaxPQ+ scheme using an optimized core-shell radius and atomic polarizability of the constituent atoms. Overall the dielectric constants for these polymer systems reasonably agree well with experiments.[22] The high dielectric constant of these systems may be attributed to the interplay between thermally activated local atomic motions that results in the enhanced fluctuation of molecular dipole moments.[22] Both PDTC-HK511 and TDI-EDR148 are rigid-base polymer containing a benzene ring and a flexible ether-amine in the backbone. This flexible segment within the backbone gives an extra ability of local chain movement thus increasing the dielectric constant, which may facilitate further polarization when subjected to an external electric field.[22]

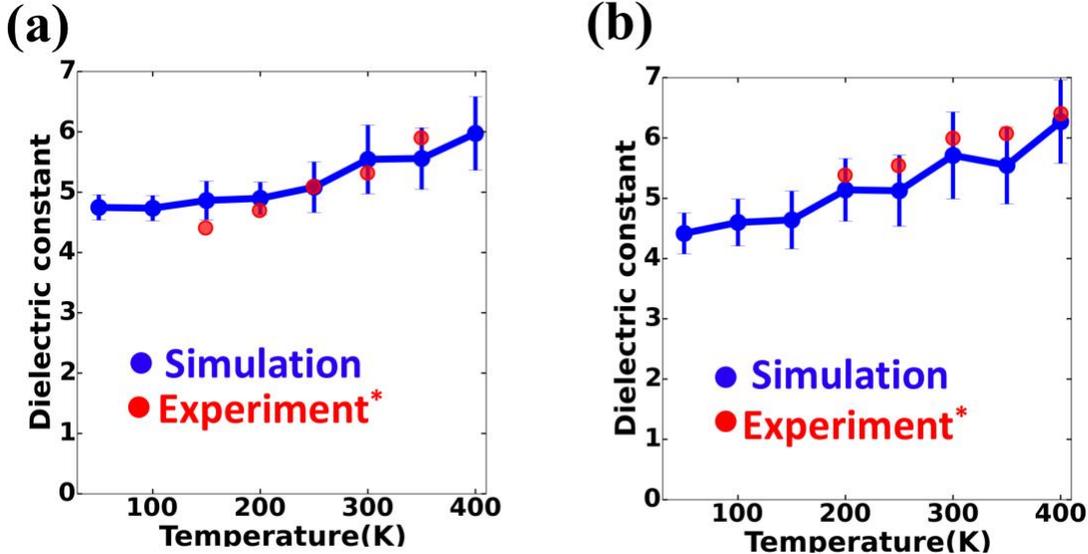

**Figure 5**: Dielectric constants of (a) PDTC-HK511 and (b) TDI-EDR148 polymer systems (blue dots) as a function of temperature. The increasing trend as well as predicted values show a good agreement with experimental values (shown in red).

### 3.2. Dielectric Behavior of DVS-BCB and FPE

DVS-BCB and FPE are promising materials that exhibit high dielectric constant values and high thermal stability.[24, 25] DVS-BCB is used in preparation of commercially available resins for multichip modules and four-level GaAs interconnect circuits.[24] DVS-BCB and FPE both have high glass transition temperature, thus allowing graceful failure and roll-to-roll production capabilities.[45] Following the previously described procedures, we have computed the temperature-dependence in dielectric constant $\epsilon_0(T)$ based on samples taken from a large number of RMD trajectories. Fig. 6 shows the temperature-dependent dielectric constant of amorphous DVS-BCB and FPE systems. Overall the obtained dielectric constants agree well with experiments.[24, 25, 46-50]

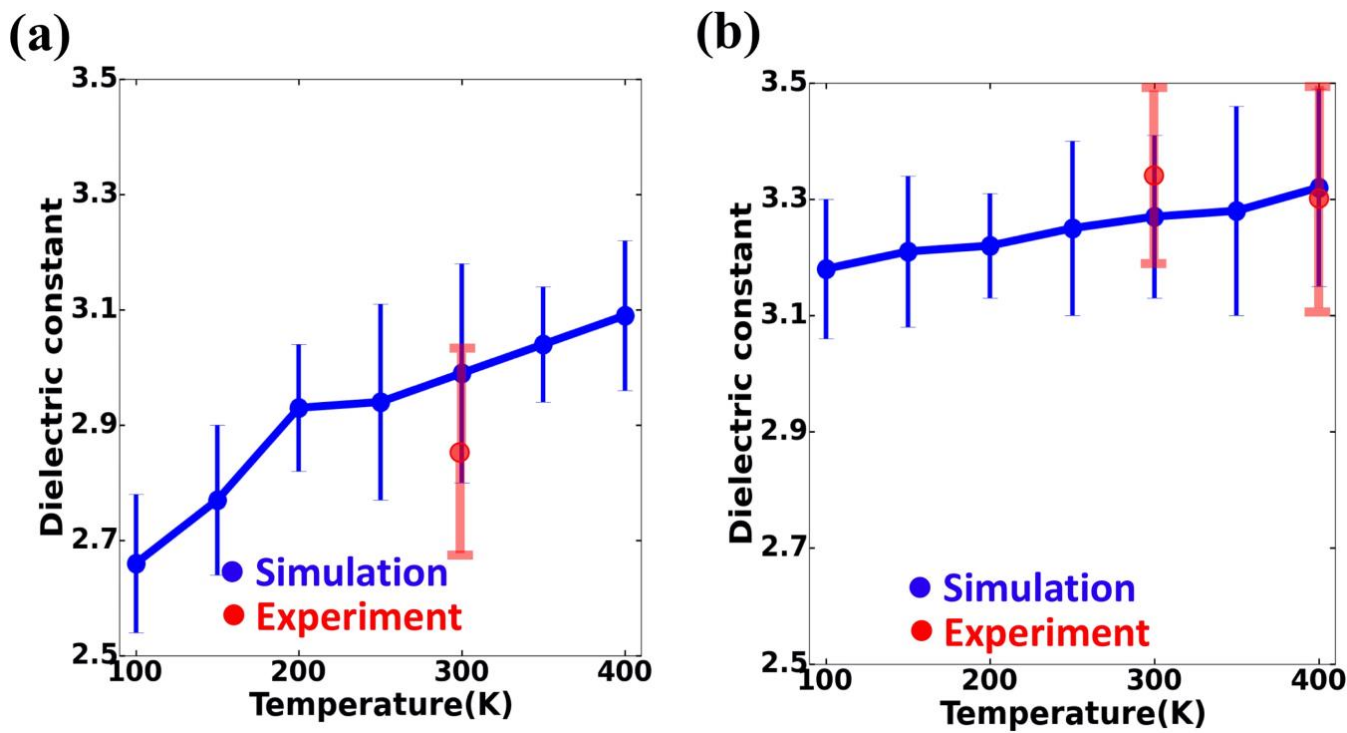

**Figure 6**: Dielectric constants of (a) DVS-BCB and (b) FPE polymer systems as a function of temperature. The predicted values show a good agreement with experiments[23, 24, 46-50].

## 4. Conclusion

In conclusion, we have developed a scalable computational framework to evaluate the temperature-dependent dielectric responses of amorphous polymer. The seamless creation of amorphous polymer structure realized by SMILES strings, Open Babel, and RXMD will allow researchers to explore the vast parameter space of high energy polymer design in a highly automated fashion. Equipped with the scalable parallel all-atom RMD simulation engine, it is possible to incorporate complex repeat units with industry-relevant chain lengths. The model parameters, such as atomic polarizability, core-shell charges and radii, are extensively optimized and validated by experiments and first-principle calculations using the high-frequency as well as temperature-dependent dielectric constants. Our computational framework has been successfully applied to the recently identified four high energy density polymers, PDTC-HK511, TDI-EDR148, DVS-BCB and FPE, and revealed that the thermally activated flexible segment motion in polymer chains facilitates the fluctuation of the molecular dipole moments. The obtained dielectric constants as well as its increasing trend quantitatively agree with the experiments. The computational framework presented here for the first time realizes high throughput theoretical screenings incorporating several key parameters in the high energy density polymer design. Therefore, combined with efficient experimental synthesis and characterization, our framework is expected to accelerate the discovery of next-generation high energy density polymer and device development.

**Acknowledgments**

This work was supported by the Office of Naval Research through a Multi-University Research Initiative (MURI) under grant number (N00014-17-1-2656).